\documentclass[12pt]{article}
\usepackage{epsfig,cite}
\def\iso#1#2{\mbox{${}^{#2}{\rm #1}$}}

\def\he#1{\iso{He}{#1}}
\def\li#1{\iso{Li}{#1}}

\def\nnu{N_\nu}

\newcommand\beq{\begin{equation}}
\newcommand\eeq{\end{equation}}
\newcommand\beqar{\begin{eqnarray}}
\newcommand\eeqar{\end{eqnarray}}
\def\la{\mathrel{\mathpalette\fun <}}

\def\fun#1#2{\lower3.6pt\vbox{\baselineskip0pt\lineskip.9pt
  \ialign{$\mathsurround=0pt#1\hfil##\hfil$\crcr#2\crcr\sim\crcr}}}

\begin{document}
\begin{titlepage}
\pagestyle{empty}
\baselineskip=21pt
\rightline{astro-ph/0408033}
\rightline{UMN--TH--2316/04}
\rightline{FTPI--MINN--04/28}
\rightline{August 2004}
\vskip 0.1in
\begin{center}
{\large{\bf New BBN limits on Physics Beyond the Standard Model from 
\he4}}
\end{center}
\begin{center}
\vskip 0.2in
{{\bf Richard H. Cyburt}$^{1}$, {\bf Brian D. Fields}$^2$
 {\bf Keith A. Olive}$^3$, and {\bf Evan Skillman}$^4$}\\
\vskip 0.1in
{\it
$^1${TRIUMF, Vancouver, BC V6T 2A3 Canada}\\
$^2${Center for Theoretical Astrophysics,
Departments of Astronomy and of Physics, \\ University of Illinois, Urbana, IL 61801, USA}\\
$^3${William I. Fine Theoretical Physics Institute, \\
University of Minnesota, Minneapolis, MN 55455, USA}\\
$^4${School of Physics and Astronomy, \\
University of Minnesota, Minneapolis, MN 55455, USA}}\\
\vskip 0.2in
{\bf Abstract}
\end{center}
\baselineskip=18pt \noindent
A recent analysis of the \he4 abundance determined from observations of 
extragalactic HII regions indicates a significantly greater uncertainty for the 
\he4 mass fraction. The derived value is now 
in line with predictions from big bang nucleosynthesis
when the baryon density determined by WMAP is assumed.  
Based on this new analysis of \he4,
we derive constraints on a host of particle properties which include:
limits on the number of relativistic species at the time of BBN (commonly
taken to be the limit on neutrino flavors), limits on the variations of 
fundamental couplings such as $\alpha_{em}$ and $G_N$, and limits on decaying
particles.

\end{titlepage}
\baselineskip=18pt

\section{Introduction}

Big bang nucleosynthesis (BBN) is one of the most sensitive available probes of
physics beyond the standard model.  The concordance between
the observation-based determinations of the light element abundances
of D, \he3, \he4, and \li7 \cite{bbn}, and their theoretically
predicted abundances reflects the overall success of the standard 
big bang cosmology.  Many departures from the standard model are
likely to upset this agreement, and  are tightly constrained \cite{mm}. 

The \he4 abundance, in particular, has often been used as a sensitive
probe of new physics.  This is due to the fact that nearly all
available neutrons at the time of BBN end up in \he4 and the
neutron-to-proton ratio is very sensitive to the competition between
the weak interaction rate and the expansion rate.  For example, a
bound on the number $g_*$ of relativistic degrees of freedom (at the time of
BBN), commonly known as the limit on neutrino flavors, $\nnu$, is
derived through its effect  on the
expansion rate, $H \propto \sqrt{g_*}$ \cite{ssg}.  However, because the
calculated \he4 abundance increases monotonically
with baryon density (parameterized by the
baryon-to-photon ratio, $\eta \equiv n_b/n_\gamma$), a meaningful limit on $\nnu$ requires
both a lower bound to $\eta$ and an upper bound to the primordial \he4
mass fraction, $Y_p$ \cite{ossty}.
Indeed, for a fixed upper limit to $Y_p$, the upper limit to $\nnu$
can be a sensitive function of the lower limit to $\eta$, particularly
if $\eta$ is small \cite{ossty,ytsso}.

The recent all-sky, high-precision measurement of microwave background
anisotropies by WMAP \cite{wmap} has opened the possibility for new
precision analyses of BBN.  Among the
cosmological parameters determined by WMAP, the baryon density has
been derived with unprecedented precision.  The WMAP best fit assuming
a varying spectral index is $\Omega_B h^2 = 0.0224 \pm 0.0009$ which is
equivalent to $\eta_{\rm 10,CMB} = 6.14 \pm 0.25$, where $\eta_{10} = 
10^{10} \eta$. This result is
sensitive mostly to WMAP alone but does include CMB
data on smaller angular scales \cite{small}, Lyman $\alpha$ forest data, and
2dF redshift survey data \cite{2df} on large angular scales. 
This result is very similar to the corresponding value obtained
from combining WMAP with SDSS data and other CMB measurements, 
which gives $\Omega_b h^2 = 0.0228^{+0.0010}_{-0.0008}$ \cite{sdss} 
and corresponds to $\eta_{10} = 6.25^{+0.27}_{-0.22}$.  Using the WMAP
data to fix the baryon density, one can make quite accurate
predictions for the light element abundances
\cite{cfo3,coc,cuoco,cyburt}. At the WMAP value for $\eta$, the \he4
abundance is predicted to be \cite{cyburt}:
\beq
Y_p = 0.2485\pm0.0005
\label{ybbn}
\eeq

On the other hand, accurate \he4 abundances have been and continue to
be difficult to obtain.  It is recognized that there are many
potential sources of systematic uncertainties in the derived \he4
abundance \cite{OSk,other}.  As a result there exists a wide range of
derived primordial \he4 abundances which have typically been
relatively low compared with (\ref{ybbn}).  Recently, a reanalysis
\cite{os2} of the \he4 data \cite{iz,iz2} has led to a significant
enlargement in the statistical uncertainty as well as a potential
shift in the mean value.  A representative result of that analysis is
\beq
Y_p = 0.2495 \pm 0.0092
\label{highhe}
\eeq
Conservatively, it would be difficult to exclude any value of
$Y_p$ inside the range 0.232 -- 0.258.

Because much of the previous work was based
on relatively low values of $Y_p$, 
tension between the value of $\eta$ inferred from either D/H or WMAP,
and \he4 gave rise to very tight constraints on $\nnu$ and
on other particle properties.  In light of the newly suggested range
for $Y_p$ \cite{os2}, it is important to reexamine the constraints on
physics beyond the standard model.  Potential limits from D/H have
been discussed recently \cite{cfo3,cuoco,cyburt}, and we will just quote those
results below in comparison with the results derived here.  At
present, it is not possible to use \li7 to obtain constraints.  This
is due to 1) the large uncertainty in the BBN prediction of the \li7
abundance, and 2) to the current discrepancy between the BBN
prediction and the observational determination of the \li7 abundance
(see e.g. \cite{cfo3,coc,cuoco,cyburt,cfo4}).

\section{The \he4 Abundance}
\label{sect:4He}

The \he4 abundance has had a somewhat checkered history over the last
decade.  Of the modern determinations, the work of Pagel et
al. \cite{pagel} established the analysis techniques that others were soon to
follow \cite{follow}.  Their value of $Y_p$ $=$ 0.228 $\pm$ 0.005 was
significantly lower than that of a
sample of 45 low metallicity HII regions, observed and
analyzed in a uniform manner\cite{iz}, with a derived value of $Y_p$ $=$ 0.244
$\pm$ 0.002.  An analysis based on the
combined available data as well as unpublished data yielded an
intermediate value of 0.238 $\pm$ 0.002 with an estimated systematic
uncertainty of 0.005 \cite{os}.  An extended data set 
including 89 HII regions obtained $Y_p$ $=$ 0.2429
$\pm$ 0.0009 \cite{iz2}.  However, the recommended value is based on
the much smaller subset of 7 HII regions, finding $Y_p$ $=$ 0.2421
$\pm$ 0.0021.  As seen in table 6 of \cite{iz2}, changing the assumed
value of the equivalent width of He absorption for one of the observed
He wavelengths by 0.1 \AA\ changes the derived abundance significantly
to $Y_p$ $=$ 0.2444 $\pm$ 0.0020.  This change of $\Delta Y_p$ $=$
0.0023, is indicative of the importance of systematic errors.
 
\he4 abundance determinations depend on a number of physical 
parameters associated with the HII region in addition to the overall
intensity of the He emission line.  These include the temperature,
electron density, optical depth and degree of underlying
absorption.  A self-consistent analysis may use
multiple  \he4 emission lines to 
determine the He abundance, the electron density and the optical
depth. In \cite{iz}, five He lines were used, underlying He absorption was assumed to be negligible 
and temperatures based on OIII observations were used.

A very accurate
helium abundance for the HII region NGC 346 in the Small Magellanic
Cloud was derived with a value of $Y_p$ $=$ 0.2345 $\pm$
0.0026 \cite{peim}.  Knowing that the OIII temperatures are systematically high,
they use the He~I emission lines to solve for the electron
temperature. Recently, the
spectra of five metal poor HII regions - NGC 346 and four regions reported in
\cite{iz} have been reanalyzed \cite{lppc}. 
After considering the effects of additional physical processes 
(e.g., collisional excitation of the Balmer lines), a higher
determination of Y$_p = 0.239 \pm 0.002$ was found.

The question of systematic uncertainties was addressed in some detail
in \cite{OSk}. It was shown that there exist severe degeneracies
inherent in the self-consistent method, particularly when the effects
of underlying absorption are taken into account.  A sixth He line was
proposed to test for the presence of underlying He absorption.
However, even in the six-line method, one can not escape the
degeneracies present in the solutions.  In particular, solutions with
no absorption and high density are often indistinguishable (i.e., in a
statistical sense they are equally well represented by the data) from
solutions with underlying absorption and a lower density.  In the
latter case, the He abundance is systematically higher.  These
degeneracies are markedly apparent when the data is analyzed using
Monte-Carlo methods which generate statistically viable
representations of the observations. When this is done, not only are the
He abundances found to be higher, but the uncertainties are also found
to be significantly larger than in a direct self-consistent approach.

In \cite{os2}, the Monte-Carlo method was applied to seven of the
highest quality observations from the sample in
\cite{iz}.  
As expected, systematically higher He abundances were
found, with significantly larger uncertainties.  The results of a
regression (to zero metallicity) led to the primordial He abundance
given in Eq. (\ref{highhe}).  With this value of $Y_p$ (particularly
with the stated uncertainty), there is clearly no discrepancy between
He observations and BBN predictions of $Y_p$ at the WMAP value for
$\eta$.  It was stressed however, that although this method found
higher He abundances, one could not exclude the lower abundances found
by other methods.

\section{Standard BBN}
\label{sect:sbbn}

Key to BBN analysis is an accurate determination of BBN
theory uncertainties, which are dominated by the errors in nuclear
cross section data.  To this end, several groups have determined
reaction rate representations and uncertainties.  Smith, Kawano and
Malaney~\cite{skm} presented the first detailed error budget for BBN,
generally assuming constant relative errors.  In more recent 
work~\cite{nb00},  uncertainties were propagated
based on available nuclear data into the light element predictions.
The NACRE collaboration presented a larger focus nuclear
compilation~\cite{nacre}, meant to update the previous astrophysical
standard \cite{cf88}.  However, their ``high'' and ``low'' limits are
not defined rigorously as 1 or 2 sigma limits (see \cite{cfo1,cvcr01}
for its impact on BBN).  In an attempt to increase the rigor of the
NACRE errors, we reanalyzed ~\cite{cfo1} the data using NACRE cross
section fits defining a ``sample variance'' which takes into account
systematic differences between data sets.

Since then, new data and techniques have become available, motivating
new compilations.  Within the last year, several new BBN compilations
have been presented~\cite{coc,cuoco,cyburt}, the latter being one of
the most rigorous and exhaustive efforts to determine reliable rate
representations and meaningful uncertainties and we adopt this compilation here.

To compare theory with observations, we will adopt the \he4 results
discussed in \S\ref{sect:4He}.  For brevity, we will simply adopt the
D observations discussed in~\cite{Dref}.  We use the D abundance
constraints based on the best five measurements of D/H in QSO
absorption line systems 
\begin{eqnarray}
\label{eq:D_hi}
{\rm D/H}_A &=& (2.78 \pm 0.29 )\times 10^{-5} 
\end{eqnarray}

\begin{figure}[ht]
\epsfig{file=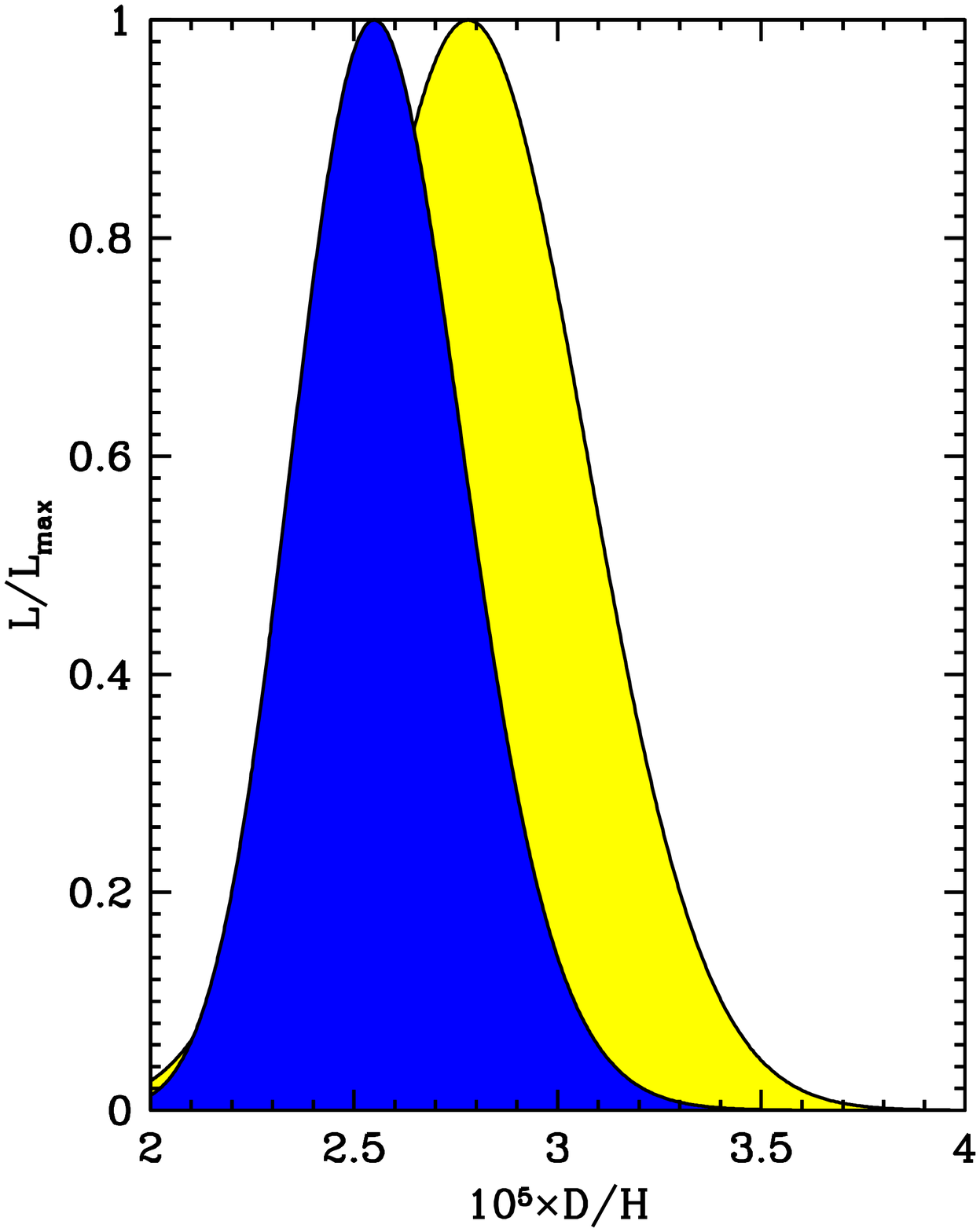,width=0.5\textwidth}
\epsfig{file=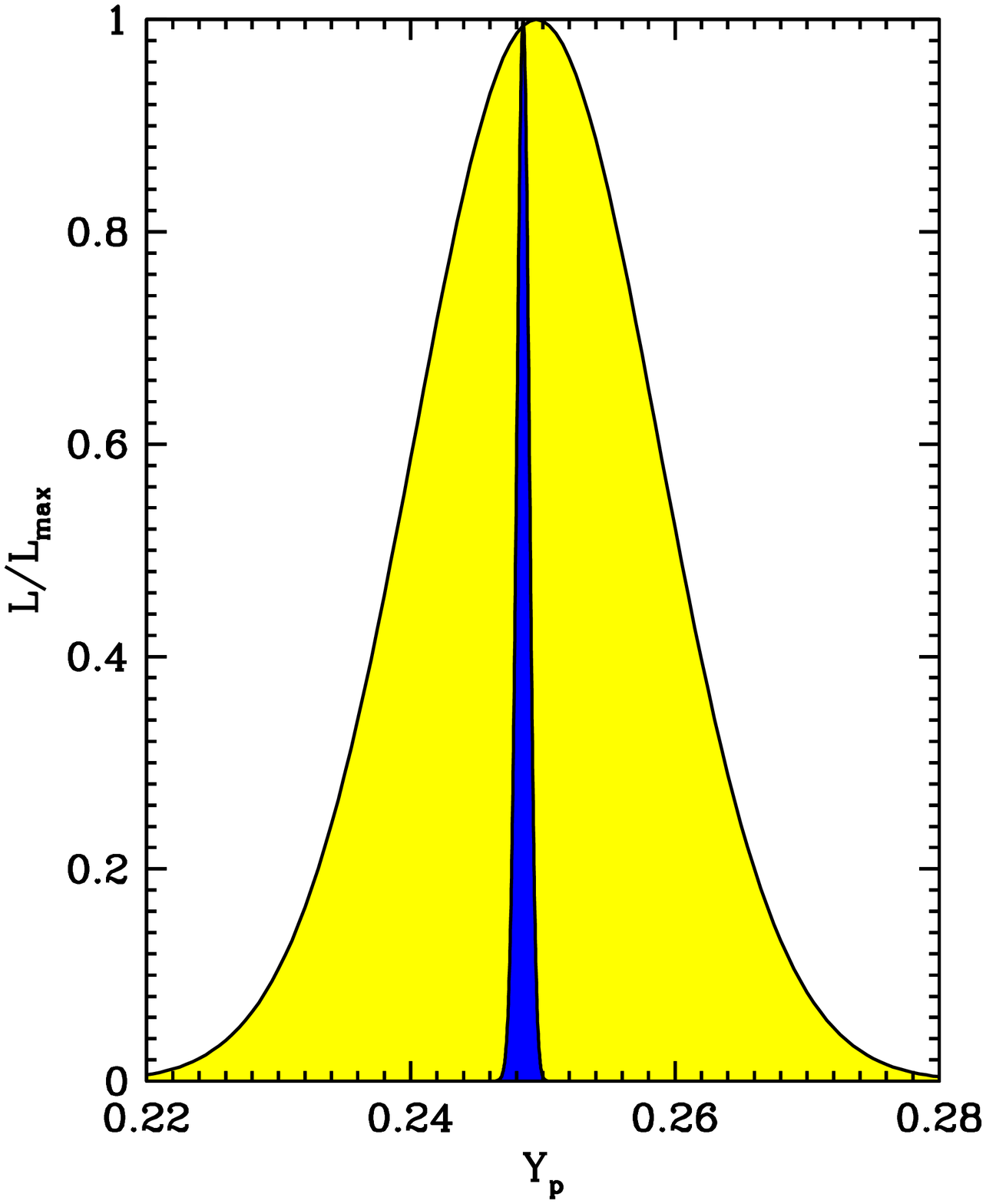,width=0.5\textwidth}
\caption{
Likelihood distributions for light element abundances.  The dark (blue) shaded
regions are the BBN predictions given the CMB-determined $\eta$
values.  The light (yellow) shaded regions correspond to the light
element observations: the new $Y_p$ and world average of D/H$_A$ values of
eq.~(\ref{eq:D_hi}) are represented by the yellow regions. 
\label{fig:likelihoods}
}
\end{figure}

BBN is tested
with the comparison between light element abundance observations, BBN
theory predictions, and the subsequent allowed ranges of baryon
density, which has now been independently measured
by CMB anisotropy experiments \cite{cfo3}.  One way to
perform this test is to use the CMB range of $\eta$ as an input to the
BBN calculation, and to compute the resulting ranges in the light
elements.  This procedure is illustrated in Figure
\ref{fig:likelihoods}.  As is well known, the agreement with D is
excellent.  The \he4 overlap is also essentially perfect, though the
larger uncertainty in \he4 renders this agreement a less powerful
test, and the near-coincidence of central values is fortuitous.
Nevertheless, the new \he4 analyses does bring this nuclide
in good agreement with the CMB and D.  This consistency was not
guaranteed and marks a success of BBN and of cosmology.

\section{Beyond the Standard Model}

For several cases of interest, it will be useful to define
a dimensionless cosmic ``speed-up'' factor $\xi = H_{\rm new}/H_{\rm
std}$, where $H = \dot{a}/a$ is the Hubble expansion rate; $\xi = 1$
then represents the unperturbed case.  The expansion rate itself is
given by the Friedmann equation, which for a flat universe is $H^2 =
(8\pi/3) \ G_N \rho$, where $\rho$ is the total mass-energy density.
Thus the speed-up factor evolves as $\xi = \sqrt{(G_N \rho)_{\rm
new}/(G_N \rho)_{\rm std}}$.  For the case of a radiation-dominated
universe, we have $\rho \propto g_* T^4$, where $g_* = 2 + 7/2+7
\nnu/4$ counts the relativistic degrees of freedom in photons, $e^\pm$
pairs, and $\nnu$ neutrino species.

\subsection{Constraints on $N_\nu$}

We first consider the canonical extension of
standard BBN, in which there are
$N_\nu$ effectively massless
($m_\nu \ll 1$ MeV) left-handed neutrino species.
The increase in the speed-up factor is 
\beq
\xi = \sqrt{1 + 7\delta \nnu/43},
\eeq
where $\delta \nnu = \nnu - 3$.
This in turn changes the weak freezeout
temperature and ultimately affects all of the
light elements.  Until recently, the effect on the
\he4 abundance was the only measurable consequence,
but D/H measurements are now sufficiently
accurate that D/H also has an important
sensitivity to $\nnu$ \cite{cfo2}.

\begin{figure}[t]
\centerline{\epsfig{file=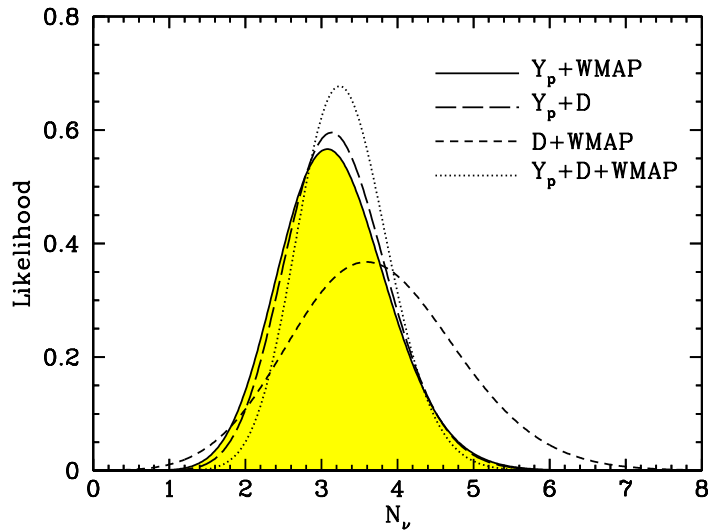,width=0.5\textwidth, angle=270}}
\caption{
The likelihood distribution for $\nnu$ based on the WMAP value of $\eta_{10} = 6.14 \pm 0.25$ and 
$Y_p$ from Eq. \ref{highhe} (shaded), WMAP and D/H (dashed), WMAP and both
$Y_p$ and D/H$_A$ (dotted).  We also show the result without the imposing the WMAP
value for $\eta$ (long dashed).
\label{fig:LN}
}
\end{figure}

For a fixed value of $\eta_{10} = 6.14 \pm 0.25$ and the He abundance given in
(\ref{highhe}), we show the likelihood distribution for $\nnu$ by the shaded region
in Fig. \ref{fig:LN}.  Also shown for comparison are the likelihood distribution
based the WMAP value of $\eta$ using D/H alone, $Y_p$ and D/H, and the
result based on BBN alone. Despite the increased uncertainty in the He abundance,
it still provides the strongest constraint on $\nnu$.  D/H is nonetheless becoming
competitive in its ability to set limits on $\nnu$. 

\begin{figure}[t]
\epsfig{file=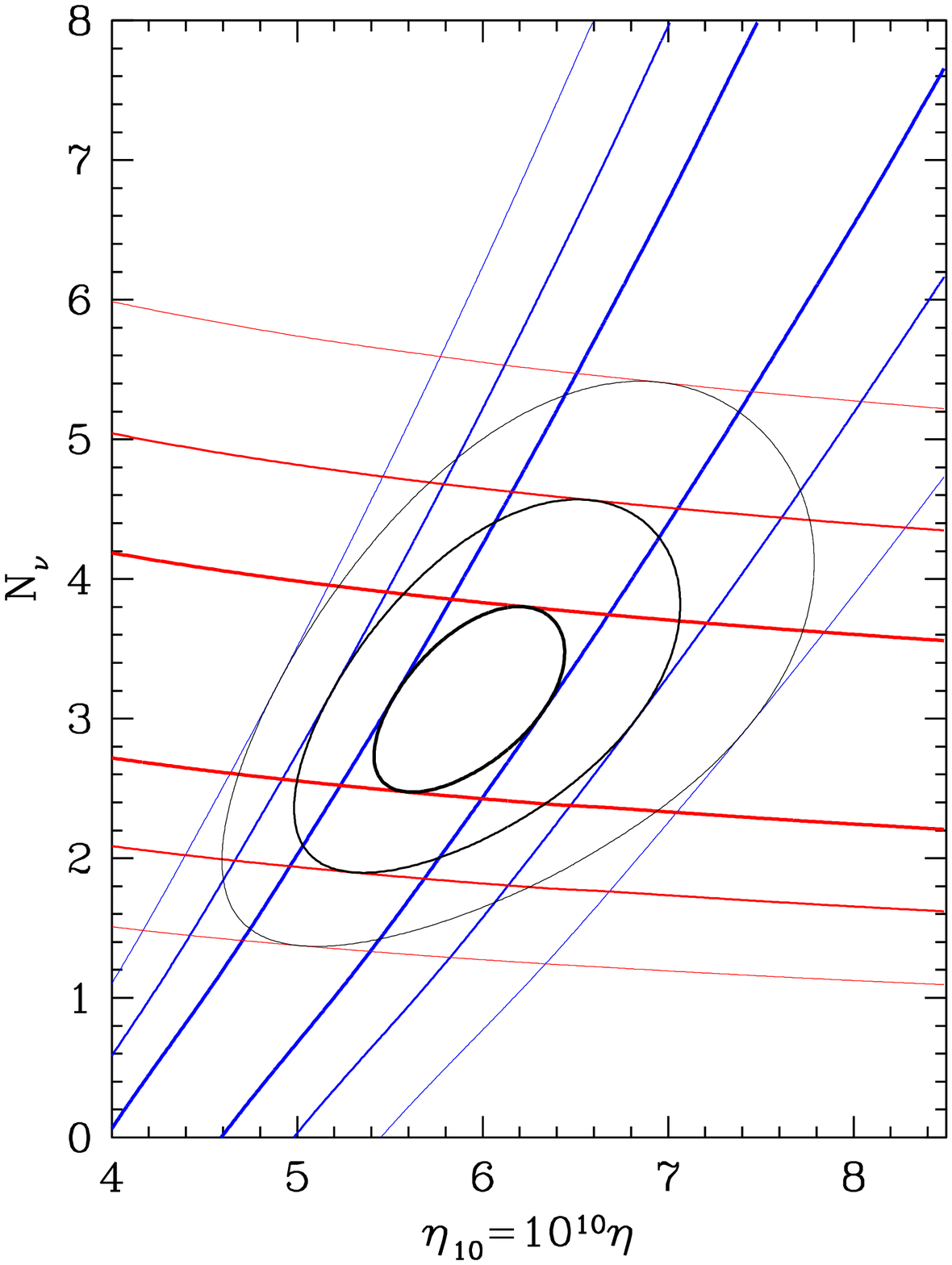,width=0.5\textwidth}
\epsfig{file=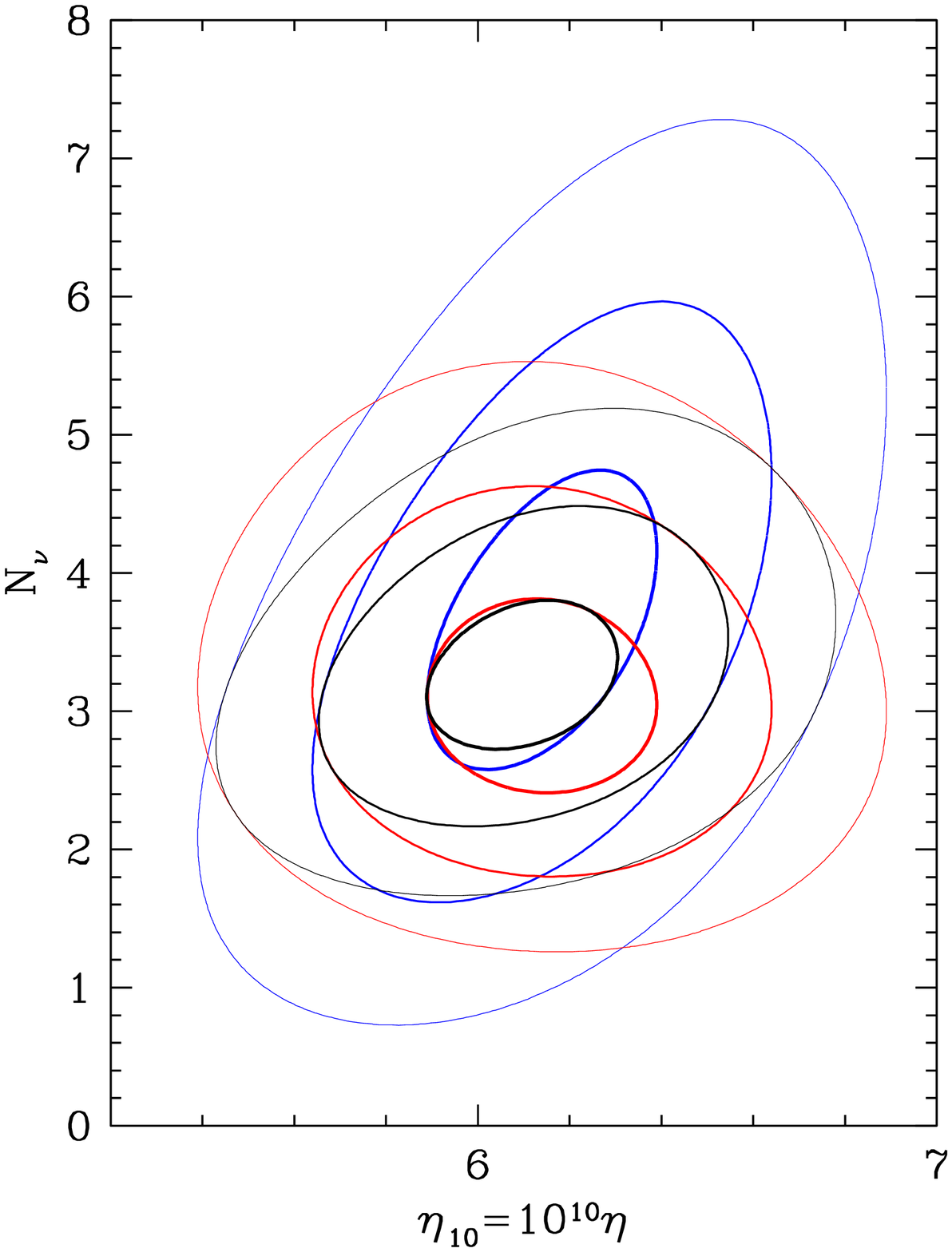,width=0.5\textwidth}
\caption{
a) BBN-only constraints on $\eta$ and $\nnu$.
The thickest (thinnest) curves correspond to $1\sigma$ ($3\sigma$)
limits.
The nearly vertical (blue) curves are limits due to D/H,
nearly horizontal (red) curves are for \he4, and 
the closed (black) contours combine both.
b) As in a), 
with the CMB $\eta$ information included.
\label{fig:L2YD}
}
\end{figure}

Figure \ref{fig:L2YD}a shows the 
joint limits on $\eta$ and $\nnu$
based on D and \he4.
We see that the \he4 contours are nearly
horizontal, which arises from the weak (logarithmic)
sensitivity of $Y_p$ to $\eta$, as opposed to a
stronger, linear sensitivity to $\delta \nnu$.
Thus \he4 by itself is a poor baryometer but
an excellent probe of nonstandard physics.
On the other hand, the D/H contours have a steep slope,
indicating a strong sensitivity to $\eta$ which
is the origin of the D/H power as a baryometer.
The non-vertical nature of the slope does however
indicate a correlation between the D/H sensitivity
to $\eta$ and $\nnu$.  Thus by combining D and \he4 we
can expect to arrive at strong constraints on both
parameters.  Numerical results appear in Table \ref{tab:barab},
where we see that these light elements alone
constrain $\eta$ to within about $10\%$, and
fix $\nnu$ to within about $20\%$, both at $1\sigma$.
Note that the contour ellipses in Figure~\ref{fig:L2YD}
have a slight positive tilt, corresponding to 
a small positive correlation between $\eta$ and $\nnu$.

\begin{table}[ht]
\begin{center}\caption{The table shows constraints placed on $N_{\nu}$ and 
$\eta$ by various combinations of observations.  Shown are the 68\%
confidence limits determined by marginalizing the 2-D likelihood
distribution ${\mathcal L}(\eta,N_{\nu})$. Also shown are the 95\%
upper limits on $\delta N_{\nu}=N_{\nu}-3$, given that $\delta N_{\nu} > 0$.}
\label{tab:barab}
\medskip
\begin{tabular}{||l|c|c|c||}
\hline\hline
Observations & \ \ $\eta_{10}\equiv 10^{10}\eta$ \ \ & $N_{\nu}$ &
$\delta N_{\nu,max}$ \\
\hline\hline
$Y_p$ + D/H$_A$ & $5.94^{+0.56}_{-0.50}$ & $3.14^{+0.70}_{-0.65}$ & 1.59 \\
\hline
$Y_p$ + $\eta_{CMB}$ & $6.14\pm0.25$ & $3.08^{+0.74}_{-0.68}$ & 1.63 \\
\hline
D/H$_A$ + $\eta_{CMB}$ & $6.16\pm0.25$ & $3.59^{+1.14}_{-1.04}$ & 2.78 \\
\hline
$Y_p$ + D/H$_A$ + $\eta_{CMB}$ & $6.10^{+0.24}_{-0.22}$ & $3.24^{+0.61}_{-0.57}$ & 1.44 \\
\hline\hline
\end{tabular}
\end{center}
\end{table}

Also appearing in Table~\ref{tab:barab}, we have shown the 95\% upper
confidence limits placed on the effective neutrino number, $\delta N_{\nu,max}$,
assuming that $N_{\nu}>3.0$ or $\delta N_{\nu}>0.0$.  
The constraints presented suggest a robust upper
bound of 1.6 with 95\% confidence.  We next introduce the CMB
information on $\eta$; this tests the overall consistency, but as we
have already shown, the agreement is good for the standard $\nnu =3$
case.  Note that CMB anisotropies also have some sensitivity to
$\nnu$, though this is at the moment significantly weaker than the
light element sensitivity.  We do not use this additional information,
which would slightly strengthen the constraints on $\nnu$, but would
not affect the $\eta$ limits (where the CMB impact is largest) due to
the independence of the CMB limits on $\eta$ and $\nnu$ \cite{gary}.

Figure \ref{fig:L2YD}b
shows the impact of the CMB on the $\eta$ and $\nnu$
constraints.  We see that the dominant effect is 
that the CMB narrows and steepens the combined
contours; this reflects the very tight CMB constraint on $\eta$.
Table \ref{tab:barab} shows the impact of the CMB on the $\eta$ and
$\nnu$ constraints.   The
resulting precision on $\eta$ is roughly doubled, to about a $4\%$ (!) 
measurement, dominated by the CMB contribution but for which the D/H
contribution is not negligible.  The precision of the $\nnu$
constraint remains essentially the same, reflecting both the dominance
of $Y_p$ in determining $\nnu$, as well as the near-independence of
$Y_p$ on $\eta$.

In \cite{os2}, it was noted that the primordial value of the \he4 abundance
based on a regression with respect to O/H was only marginally
statistically more significant that a weighted mean which yields
$Y_p = 0.252 \pm 0.003$.  This result is also obtained 
using a Bayesian analysis in which the sole prior is
the increase in \he4 in time \cite{hos}. The combination of 
the \he4 abundance based on the mean value and the CMB value for
$\eta$ gives $\nnu = 3.27 \pm 0.24$ with a 95\% upper limit $\delta N_{\nu,max}$ = 0.7.
Recall, the constraint on $\delta N_{\nu,max}$ assumes $N_\nu>3$ or $\delta N_\nu>0$.

In all cases the preferred values
for $\nnu$ are consistent with $\nnu =3$,
and in many cases are much closer to $\nnu = 3$ than
$1\sigma$.  This restates the overall consistency
among standard BBN theory, D and \he4 observations,
and CMB anisotropies. It also constrains departures
from this scenario.
Our combined limit using BBN + light elements + CMB limit is:
\beq
\label{eq:Nnu}
2.67 \le \nnu \le 3.85 
\eeq
at 68\% CL.

\subsection{Constraints on the Variation of Fundamental Constants}

As noted earlier, BBN also placed interesting limits on possible variations
of fundamental constants.  Indeed, almost every fundamental parameter
can be constrained by BBN if it affects either the expansion rate of the Universe,
the weak interaction rates prior to nucleosynthesis, or of course the nuclear rates
themselves. 
 As quantitative examples of the constraints which can be derived,
we focus here solely on variations of Newton's constant and the fine-structure
constant. Here, we simply note that many other constraints have been considered
in the past (for a recent review see: \cite{uzan}).

\subsubsection{Newton's Constant, $G_N$}

Strictly speaking, it makes no sense to consider the variation of 
a dimensionful constant such as $G_N$ (see e.g. \cite{duff}).
We include the discussion here for the purposes of comparison 
with previous constraints.  We also note that the limit set here could
be interpreted as a limit on the gravitational coupling 
between two protons ($G_N m_p^2$) in a framework
where we have chosen $m_p$ (and all other particle masses) to be constant.
Early constraints \cite{earlyG} on $G_N$ relied mainly on \he4
abundance observations. Recently, the D/H abundance was used
\cite{cdk} in conjunction with the WMAP determination of $\eta$ to set
a limit on $\Delta G_N / G_{N,0}$ of about 20\% from the time of
BBN. Assuming a simple power law dependence $G_N \sim t^{-x}$, $x$ was
constrained to the range $-0.004 < x < 0.005$ implying $ -4 \times
10^{-13}$yr$^{-1} < {\dot G_{N,0}} / G_{N,0} < 3 \times
10^{-13}$yr$^{-1}$ 
The use of D/H was motivated in
part by the previously discrepant value for $Y_p$.

Although the uncertainty in the \he4 abundance has been argued to be
significantly larger than past values (0.009 vs 0.002) \cite{os2}, the
resulting bounds on $G_N$ are still interesting.  The limit on $\delta
\nnu$ of $-0.60 < \delta \nnu < 0.82$ from the $Y_p$ + CMB constraint, 
can be translated directly to a bound on the speed-up factor: $0.949 <
\xi < 1.062$.  Any variation in $G_N$ can be expressed through $\xi$
\beq
-0.10 < {\Delta G_N \over G_{N,0}} = \xi^2 - 1 < 0.13
\eeq
or perhaps more simply put, the limit on the variation in $G_N$
can always be related directly to the limit on $\nnu$ through
\beq
{\Delta G_N \over G_{N,0}} = {7 \over 43} \delta \nnu
\eeq

If one makes the common assumption that $G_N \sim t^{-x}$, we obtain,
$-0.0029 < x < 0.0032$ and $ -2.4 \times 10^{-13}$yr$^{-1} < {\dot
G_{N,0}} / G_{N,0} < 2.1 \times 10^{-13}$yr$^{-1}$ (using
$t_0=13.7\pm0.2$ Gyr~\cite{wmap},  and  $t_{BBN} \sim 100$
sec).    Thus despite the increased
uncertainty in $Y_p$, the \he4 abundance still provides the strongest
possible constraint on $G_N$.

\subsubsection{The Fine-Structure Constant, $\alpha$}

There has been a great deal of activity surrounding 
possible variations of the fine-structure constant, motivated
largely by a reported observational analysis
of quasar absorption systems which has been interpreted as
a variation in $\alpha$ \cite{murphy3}. We note that 
other observations using similar methods \cite{chand} have not
confirmed the variation in $\alpha$, and other 
interpretations based on the nucleosynthesis of heavy Mg isotopes
in the absorbers may also explain the data \cite{amo}.
Other constraints from the CMB \cite{cmb}, the Oklo reactor \cite{Oklo,opqccv},
and meteoritic abundances \cite{opqccv,fi} have also been derived.
Once again, our goal here is to update the BBN bound on
variations in $\alpha$ using the newly derived value of $Y_p$.

If we assume that only $\alpha$ is allowed to vary (i.e., we assume
that all other fundamental parameters are held fixed), the dominant 
contribution to a change in $Y_p$ comes from the variation in 
the neutron-proton mass difference, $Q=m_n-m_p$ \cite{kpw}.  
The \he4 abundance can be estimated simply from the ratio of the
neutron-to-proton number densities, $n/p$, by assuming that essentially
all free neutrons are incorporated into \he4.  
The neutron-to-proton ratio at weak freezeout is
$(n/p)_f \sim e^{-Q/ T_f}$,
modulo free neutron decay, where $T_f $
is the temperature at which the weak interaction rate for interconverting neutrons 
and protons falls below the expansion
rate of the Universe. 

Variations in $Q$ leads to a variation in $Y_p$
given approximately by
\beq
{\Delta Y \over Y} \approx
 - {\Delta Q \over Q} 
\label{dy}
\eeq
One can write the nucleon mass difference as 
\beq
Q \sim a \alpha \Lambda_{QCD} + b v \label{dm}
\eeq
where $a$ and $b$ are dimensionless constants giving the relative
contributions from the electromagnetic and weak interactions.  In
(\ref{dm}), $v$ is the standard model Higgs expectation value.  A
discussion on the contributions to $Q$ can be found in \cite{pn}.  The
constants $a$ and $b$ are chosen so that at present the two terms
contribute -0.8 MeV and 2.1 MeV respectively.  Eqs. (\ref{dy}) and
(\ref{dm}) can be combined to give
\beq
{\Delta Y \over Y} \approx     0.6 {\Delta \alpha \over \alpha}
\eeq
Thus the current uncertainty in the observational determined value of $Y_p$
leads to a bound of $|\Delta \alpha / \alpha| < 0.06$.
If changes in $\alpha$ are correlated to changes in other gauge or Yukawa
couplings, this limit improves (in a model dependent way) by about 
2 orders of magnitude \cite{co}.

\subsection{Limits on Decaying Particles}

An exotic scenario often considered is that of late-decaying particles
($\tau_X\sim 10^8$ sec)~\cite{cefo}.  The particles are assumed to
decay electromagnetically, meaning that the decays inject
electromagnetic radiation into the early universe. If the decaying
particle is abundant enough or massive enough, the injection of electromagnetic
radiation can photo-erode the light elements created during primordial
nucleosynthesis.  The theories we have in mind are generally supersymmetric, in
which the gravitino and neutralino are the next-to-lightest and
lightest supersymmetric particles, respectively, but the constraints
hold for any decay producing electromagnetic radiation.  We thus
constrain the abundance of such a particle given its mean lifetime
$\tau_X$.  The abundance is constrained through the parameter
$\zeta_X\equiv 2E_{inj}n_X/n_\gamma$.  We can see there is a degeneracy
between the relative abundance $n_X/n_\gamma$ and the injected energy
$E_{inj}$.  Specific theories can relate the lifetime with the mass of
the decaying particle and thus the injected energy $E_{inj}$, however,
we will restrict ourselves to the general case.

The constraint placed by the \he4 abundance comes from its lower
limit, as this scenario destroys \he4.  The maximum value our
abundance parameter can take is proportional to the \he4 constraint,
$\zeta_{X,max}(\he4) \propto (Y_{BBN}-Y_{min})/Y_{BBN}$.  Here
$Y_{BBN}$ is the predicted \he4 abundance given $\eta$ and $Y_{min}$
is the minimum allowed value from observations. Using $Y_{min} = 0.232$,
we find:
\beq
\label{eqn:cefo}
\zeta_X(\he4) < 2.1\times 10^{-10}\ {\rm GeV} \left(\frac{\eta_{10}}{6.14}\right) \left(\frac{\tau_X}{10^{8} {\rm sec}}\right)^{1/4} \ \ \mbox{\rm for $\tau_X > 10^8$ sec}
\eeq
However, in this scenario the limits on the tertiary production of \li6 
provide stronger constraints on late-decaying particles, with $\zeta_X
< 5.1\times 10^{-12}$ GeV following the same scalings with $\eta_{10}$
and $\tau_X$ as \he4.

\subsection{Other Bounds}

Models of neutrino masses with new right-handed interactions
are also subject to the constraint from \he4. If we
assume right-handed neutrinos are present and
are light enough ($\la 1$ MeV) 
to count as extra relativistic degrees of
freedom during BBN, they are  subject to the constraint on $\delta
N_{\nu} < 1.60$. 
Assuming there are 3 standard model neutrinos and 3 right-handed neutrinos, 
we can relate
the limit on $\delta N_{\nu}$ to the temperature ratio of right-handed
to left-handed neutrino temperatures: 
$\delta N_{\nu} = 3(T_{\nu_R}/T_{\nu_L})^4$ \cite{oss}.  
We find $T_{\nu_R}/T_{\nu_L} < 0.85$ with
95\% confidence.  In order to change this ratio from unity,
right-handed neutrinos must decouple before epochs of entropy release,
generally due to particle annihilation.  To accommodate the constraint
on the temperature ratio, $\nu_{\rm R}$ must decouple at least before the
annihilation epoch of pions, if not before the quark-hadron
transition, with $T_{\rm dec}>140$ MeV.

As a consequence, right-handed interactions governed by some
new mass scale $M_R$,  can also be constrained.  
Requiring that these interactions decouple before the
pion annihilation epoch and recalling that $T_{dec}$ scales as
$M_R^4$, we find that:
\beq
M_{W_R} > 3.3\ {\rm TeV} \left(\frac{T_{\rm dec}}{140\ {\rm MeV}}\right)^{3/4}
\eeq
Note this constraint is stronger than limits
currently set by high-precision electroweak constraints ($M_{W_R}>715$
GeV~\cite{pdg}).  
For recent analyses of models of this type, see \cite{lang}.

BBN can also constrain the presence of vaccum energy
during nucleosynthesis, via the expansion speedup
provided by the additional energy density.
The presence of vaccuum energy during BBN
is motivated both by considerations of
dark energy\cite{fj,bs} as well as inflation.
The new vacuum energy component is typically
modelled as some scalar field $\phi$.  The evolution
of the scalar field is controlled by its
potential as well as the Friedmann equation.
However, for an important class of models,
there is a ``tracker'' solution which is
an attractor, and leads the vaccum energy density to
be proportional to the dominant radiation density
\cite{fj}.
In this case, the vacuum energy acts essentially
as an extra neutrino species, with 
\beq
\Omega_\phi(1 \ {\rm MeV}) 
 = \frac{\rho_{\rm \phi}}{\rho_{\rm rad}+\rho_\phi}
 = \frac{7 \delta \nnu/4}{10.75+7 \delta \nnu/4}
\eeq
Using our limit $\delta \nnu < 1.60$, we find that
$\Omega_\phi(1 \ {\rm MeV}) < 0.21$.  For the case
of an exponential potential 
$V(\phi) = M_{\rm Pl}^4 e^{- \lambda \phi/M_{\rm Pl}}$,
this constraints the coupling 
$\lambda = 2/\sqrt{\Omega_\phi} > 4.4$.

\section{Summary}

A new and detailed assessment \cite{os2} 
of the observed primordial
\he4 abundance, and its uncertainties, has
important implications for cosmology and BBN generally, and
for early universe and particle physics in particular.
The observed \he4 abundance is now found 
to be consistent with the $\eta$ value
given by D, leaving \li7 alone in discordance.
Moreover, both D and now \he4 are consistent
with the precision $\eta$ range determined
by recent observations of CMB anisotropies.
The newfound \he4 agreement arises primarily because
the new and more detailed error budget is also larger
than previous estimates.  Nevertheless, with
\he4 in concordance, it now rejoins D and the CMB
as a probe of physics beyond the Standard Model.

We have surveyed the impact of the new \he4 analysis on nonstandard 
physics.
The classic neutrino counting argument
is found to give relaxed though non-trivial limits to
$\delta \nnu$ 
(eq.~\ref{eq:Nnu} and Table \ref{tab:barab}).
We find that D and \he4 observations 
each make similar contributions to these limits.
Thus, if one of these observations can
be signficantly improved, it will
dominate our ability to probe exotica.

These limits immediately translate to
tighter constraints on variation of
the fundamental constants.
They also constrain the presence of
vaccuum energy (in the ``tracker'' regime)
during BBN.  Finally, we note that the
new \he4 limit on decaying particle sceanrios remains
weaker than the constraints placed by \li6;
with a factor $\sim 2$ improvement in precision,
\he4 would be competetive.

We thus urge continued effort to (realistically!)
improve the precision of the observations of
primordial abundances.  A sharpening in either
D or \he4 will immediately tighten 
the limits we place here.
And of course, the outstanding problem with
\li7 continues to demand explanation.

\section*{Acknowledgments}

R.H.C. would like to thank C. Angulo, P. Descouvemont and P. Serpico for helpful discussions on our respective BBN rate compilations~\cite{coc,cuoco,cyburt}.
The work of K.A.O. was partially supported by DOE grant
DE-FG02-94ER-40823. The work of B.D.F. was supported by the National
Science Foundation under grant AST-0092939.  The work of R.H.C. was
supported by the Natural Sciences and Engineering Research Council of
Canada.

\end{document}